\documentclass[10pt,aps,prl,showpacs,nofootinbib,preprintnumbers,a4paper,twocolumn,superscriptaddress]{revtex4-1}
\pdfoutput=1
\usepackage{amsmath}
\usepackage{amssymb}
\usepackage[pdftex]{graphicx}
\usepackage[english]{babel}
\usepackage{amsfonts}
\usepackage{bbold}
\usepackage{bm}
\usepackage{color}
\usepackage{xcolor}
\usepackage[colorlinks]{hyperref}
\usepackage{cancel}
\usepackage{units}
\usepackage{feynmf}
\hypersetup{
pdfauthor={},
pdftitle={Electroweak symmetry breaking by QCD}
}

\newcommand{\SU}[1]{\ensuremath{\mathrm{SU}(#1)}}

\begin{document}
\allowdisplaybreaks[1]
\widetext
%%%%%%%%%%%%%%%%%%
%%  TITLE PAGE  %%
%%%%%%%%%%%%%%%%%%

\title{Electroweak Symmetry Breaking via QCD}
\vspace*{10mm}
\author{Jisuke Kubo}
\email{jik@hep.s.kanazawa-u.ac.jp}
\affiliation{Institute for Theoretical Physics, Kanazawa University, Kanazawa 920-1192, Japan}

\author{Kher Sham Lim}
\email{khersham.lim@mpi-hd.mpg.de}

\author{Manfred Lindner}
\email{lindner@mpi-hd.mpg.de}
\affiliation{Max-Planck-Institut f\"{u}r Kernphysik, Saupfercheckweg 1, 69117 Heidelberg, Germany}

\begin{abstract}
We propose a new mechanism to generate the electroweak scale
within the framework of QCD, which is extended to include conformally invariant
scalar degrees of freedom belonging to a larger irreducible representation of
$\SU3_c$. 
The electroweak symmetry breaking is triggered dynamically via the Higgs portal by
the condensation  of the colored scalar field  around $1$ TeV.
The mass of the colored boson is restricted to be $350$ GeV $\lesssim m_S\lesssim 3$ TeV, with the upper bound obtained from perturbative renormalization group evolution.
This implies that the colored  boson can be produced  at LHC.
If the colored boson is electrically charged, 
the branching fraction of the Higgs decaying into two photons
can slightly increase, and moreover, it can be produced at 
future linear colliders. Our idea of non-perturbative EW scale generation can serve as a new starting point for more realistic model building in solving the hierarchy problem.
\end{abstract}

\maketitle

\section{Introduction}
With only the Standard Model (SM) Higgs like particle discovered and no new particle beyond the SM being found, there is no evidence for any of the generally proposed solutions to the hierarchy problem. With the current measured Higgs mass and the top quark mass, the SM could even survive up to the Planck scale \cite{Holthausen:2011aa,*Degrassi:2012ry}. However, one has to face the puzzle of why the electroweak (EW) scale is many orders of magnitude smaller than the Planck scale. A possible solution for the hierarchy problem is based on scale invariance, which is violated at the quantum level
and hence a scale is introduced:
The EW scale is generated dynamically by either the Coleman-Weinberg mechanism \cite{Meissner:2006zh,*Foot:2007as,*Foot:2007iy,*Holthausen:2009uc,*Iso:2009ss,*Englert:2013gz,*Farzinnia:2013pga,*Dermisek:2013pta,*Carone:2013wla,*Abel:2013mya,*Radovcic:2014rea,*Hill:2014mqa} or dimensional transmutation of a non-perturbatively created scale in a strongly coupled hidden sector \cite{Hambye:2007vf,*Hur:2007uz,*Hur:2011sv,*Holthausen:2013ota,*Heikinheimo:2013fta,*Hambye:2013dgv}.
Many of these attempts to generate the EW scale radiatively rely on the Higgs portal $\lambda_{HS} S^\dagger S H^\dagger H$,
%\begin{align}
%\lambda_{HS} S^\dagger S H^\dagger H,
%\label{eq:gghh}
%\end{align}
where the additional scalar field $S$ (charged or neutral under a certain gauge group) obtains a vacuum expectation value (VEV) either directly or indirectly. 
In this letter we propose a new non-perturbative mechanism to generate the EW symmetry breaking (EWSB) scale. Though the hierarchy problem between the EW scale and the Planck scale is not
completely solved in our proposed minimal model,
which is a least extension of the SM, 
our mechanism  can be applied to more realistic model building scenario in solving the hierarchy problem. Specifically EWSB is triggered by the condensation of an additional scalar field $S$,
which belongs to a larger representation
of $\SU3_c$.
In general the condensation of $S$, i.e. $\langle  S^\dag S\rangle \neq 0$, takes place when
\begin{align}
C_2(S)\alpha(\Lambda)\gtrsim 1,
\label{eq:c2as}
\end{align}
with $C_2$ representing the quadratic Casimir operator of a certain representation $\bm{R}$ of $S$ and $\alpha$ is the gauge coupling of the chosen non-abelian gauge group. The crucial point to notice here is that confinement (we throughout assume that the confinement scale is the same as the condensation scale) can take place even if $\alpha$ is relatively small, provided that the representation of $S$ is large enough. 
QCD is a part of the SM and generates
dynamically an energy scale of $\mathcal{O}(\unit[1]{GeV})$ by the gluon and quark condensates.
However, we emphasize that these scales are 
closely related to the fact that the quarks belong to the fundamental 
representation of $\SU3_c$.
Therefore, according to Eq.~\eqref{eq:c2as}, if  
there exist colored degrees of freedom 
belonging to a  larger  representation of $\SU3_c$,
QCD can generate much  higher energy scale in principle.
In fact exotic quarks that are confined at higher energy scale have been considered in Refs.~\cite{Marciano:1980zf,*Lust:1985aw}. However most of these exotic fermions with EW charges cannot generate the correct EW scale without large deviations from EW precision tests.
This situation will change if we consider a colored EW singlet scalar field,
as we will see in the next sections.

\section{Electroweak symmetry breaking by scalar QCD}
We assume that the SM with the new scalar QCD extension is classically scale invariant and the EW scale is generated via the condensation scale of $S$. In fact, the $\SU3_c$ sector of the SM itself before EWSB is scale invariant, contrary to ordinary QCD with explicit massive quarks. The full Lagrangian is given as
\begin{align}
\mathcal{L}=&\mathcal{L}_{\mathrm{SM},m^2\rightarrow0}+(D_{\mu,ij}S_j)^\dagger(D^\mu_{ik}S_k) \nonumber \\
&+\lambda_{HS}H^\dagger H S^\dagger S-\lambda_{\bm{1}_i}\left[\bar{\bm{S}}\times \bm{S}\times \bar{\bm{S}} \times \bm{S}\right]_{\bm{1}_i},
\label{eq:lagrangian}
\end{align}
where $D^\mu_{ij}=\delta_{ij}\partial^\mu -ig_s (T_R)_{ij}^kG^\mu_k$ and $T_R$ represents the generator for the representation $\bm{R}$ of $\SU3_c$. The term $\lambda_{\bm{1}_i}$ denotes the quartic scalar coupling for the $i$-th invariant formed by the four tensor products of the $S$ representation. Due to classical scale invariance, the Lagrangian in Eq.~\eqref{eq:lagrangian} does not contain quadratic and cubic terms of $S$. Conventional scalar QCD would be quadratically sensitive to an embedding scale and it would therefore not solve the hierarchy problem. Note, however, that our scenario is based on conformal QCD which should have only logarithmic scale dependence. Note that accidental $\mathrm{U}(1)$ symmetry appears for the $S$ sector due to the absence of cubic term and this has interesting phenomenology on its own if this $\mathrm{U}(1)$ is identified with the $\mathrm{U}(1)_Y$ hypercharge of the SM, which we will discuss later. EWSB triggered by QCD is as follows: The strong coupling $g_s$ runs as 
usual from a finite value set at high energy (Planck or GUT) scale to the condensation scale of $S$. The scalar condensate $\langle S^\dagger S \rangle$ forms when Eq.~\eqref{eq:c2as} is satisfied for $\mathcal{O}(\unit[1]{TeV})$ where the small 
value of $\alpha_s(\Lambda=\unit[1]{TeV})\approx 0.09$ is compensated by the large $C_2$ of $S$ in higher representation. Note that the confinement scale is fixed once a representation for $S$ is chosen, see Table \ref{tab:representation}. The condensate generates a scale which enters the portal 
\begin{align}
\lambda_{HS}\langle S^\dagger S \rangle H^\dagger H \rightarrow \lambda_{HS}\Lambda^2 H^\dagger H, 
\end{align}
and triggers the EWSB radiatively. The Higgs mass after EWSB is determined by 
\begin{align}
m_h^2=2 \lambda_{HS} \Lambda^2, 
\end{align}
and this in turn determines the value of Higgs quartic coupling $\lambda_h =\lambda_{HS}\Lambda^2 / v^2$,
%\begin{align}
%\end{align}
with the Higgs VEV $v=\unit[246]{GeV}$. The coupling $\lambda_{HS}$ is determined once the confinement scale is fixed to be any value higher than the EW scale, as we require that confinement happens before EWSB. In general we have no upper bound on $\Lambda$, except that larger representation of $S$ is required as $\alpha_s$ decreases with higher value of $\Lambda$. 

The low energy QCD remains unaltered by our new additional field as the coupling of higher representation of field $S$ with the quarks in fundamental representation to form a singlet requires typically higher dimensional operators. It is important to remember that such condensation takes place albeit the small coupling of $\alpha_s$ at scales of $\mathcal{O}(\mathrm{TeV})$ due to a large $C_2$ value for larger representation.
As we can read off from Table \ref{tab:representation}, $\bm{15'}$ is the unique representation for our purpose as it generates the desired condensation scale at $\mathcal{O}(\unit[1]{TeV})$. 
%The smaller representations can not generate a correct Higgs mass $m_h$ as the coupling $\lambda_{HS}$ has to be larger than $20$  even for the next possibility $\bm{10}$, which is clearly out of our consideration. 

\begin{table}
 \begin{tabular}{|c|c|c|c|}
 \hline
  Rep $(\bm{R})$& $C_2(\bm{R})$ & $C(\bm{R})$ & $\Lambda$ (GeV)\\ \hline
  $\bm{8}$ & $3$ & $3$ & $1$ \\
  $\bm{10}$ & $6$ & $15/2$ &$20$ \\
  $\bm{15}$ & $16/3$ & $10$ & $10$ \\
  $\bm{15'}$ & $28/3$ & $35/2$ & $1000$\\
  $\bm{21}$ & $40/3$ & $35$ & $10^5$\\ \hline
 \end{tabular}
\caption{Values of the quadratic Casimir and index for certain representations of QCD. The approximate confinement scale $\Lambda$ for each representation is listed.}  
\label{tab:representation}
\end{table}

%\begin{figure}
%\includegraphics[width=0.45\textwidth]{running.pdf}
%\caption{The running of scalar couplings with some generic boundary conditions at the scale of $\unit[1]{TeV}$.\label{fig:running}}
%\end{figure}

The phenomenology of this new scalar QCD extension with the representation of $S$ being $\bm{15'}$ will now be discussed in detail. First we can constrain the coupling $\lambda_{\bm{1}_i}$ and $\lambda_{HS}$ from the requirement that all the scalar couplings do not hit a Landau pole or destabilize the vacuum. For the case of $\bm{15'}$, we have 3 quartic couplings $\lambda_{\bm{1}_i}$ due to the existence of 3 invariants formed from the four tensor products of $\bm{15'}$. The invariants formed by the tensor products can be calculated with proper Clebsch-Gordan coefficients and subsequently one-loop beta functions for the quartic couplings can be calculated \cite{Cheng:1973nv}. To simplify our calculation further we assume that the order of each $\lambda_{\bm{1}_i}$ is roughly the same, i.e. $\lambda_{\bm{1}_i}\approx \lambda_S/3$ and normalized such that the mass term $m_S$ of $S$ can be extracted from the Lagrangian. Notice that the bare $m_S$ of $S$ does not exist in Eq.~\eqref{eq:lagrangian} due 
to scale invariance. The mass term can be approximately obtained from self-consistent mean field approximation \cite{Kunihiro:1983ej} after confinement has taken place, where the mean field serves as a back-reaction to the field $S$ and the mass is obtained from
\begin{align}
\frac{\lambda_S}{2}(S^\dagger S)(S^\dagger S) \rightarrow \lambda_S \langle S^\dagger S \rangle S^\dagger S= \lambda_S \Lambda^2 S^\dagger S. 
\end{align}
The coupling $\lambda_S$ dictates directly $m_S^2=\lambda_S \Lambda^2$ while the mixing parameter $\lambda_{HS}$ determines $m_h$. The large $m_S$ prevents the $S$ field from obtaining non-zero VEV, hence color symmetry is not spontaneously broken. From the renormalization group equation (RGE) analysis we obtain the running of scalar couplings once the confinement scale is set. The measured $m_h$ fixes $\lambda_{HS}$, while
%The result for certain generic values of $\lambda_S$ with $\Lambda=\unit[1]{TeV}$ is shown in Fig.\ref{fig:running}. 
%The coupling $\lambda_{HS}$ is fixed once we have set the confinement scale, as they are related by the measured $m_h$. 
the mass $m_S \sim \lambda_S$ cannot be pushed arbitrarily high due to the emergence of Landau pole, yielding an upper bound on $m_S \lesssim \unit[3]{TeV}$ while the lower bound can be obtained from the collider phenomenology. The running of scalar mixing parameter $\lambda_{HS}$ is relatively slow and it will only hit the 
triviality bound when $\lambda_S$ hits the Landau pole, this subsequently drives $\lambda_h$ to a Landau pole. We would like to stress that other RGE scenarios maybe viable if the parameters $\lambda_{\bm{1}_i}$ and the confinement scale $\Lambda$ are varied independently. In this letter we study only the simplest model to explain EWSB triggered by QCD. More realistic models should include dark matter and neutrino masses and their coupling to our new scalar could alter the high UV behaviour of the RGE of λS significantly. The Landau pole at 10 TeV may therefore be absent in a more realistic model, or be a signal for non-perturbativity.

%We would expect a GUT type or new non-perturbative UV physics appearing at low energy, possibly at $\unit[10]{TeV}$, to reconcile the Landau pole problem. We will leave the UV completion issue for future investigation, but we stress that the requirement of low UV scale implies that our model can be tested or ruled out at the LHC.

\section{Collider Phenomenology}
\begin{figure}
\includegraphics[width=0.4\textwidth]{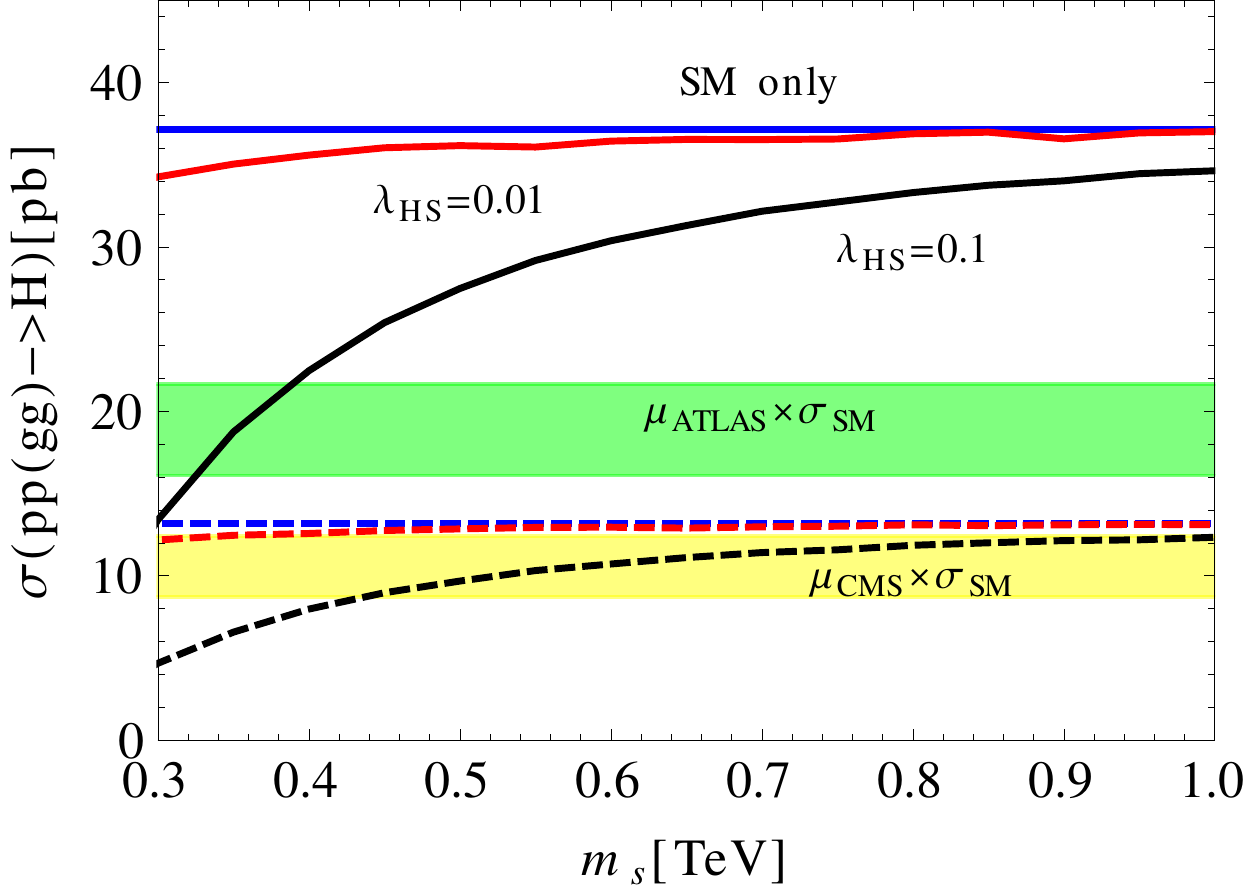}
\caption{The Higgs production cross section from gluon fusion channel at NLO is calculated for different values of $\lambda_{HS}$. The solid (dashed) curves represent the prediction of $\sigma(gg\rightarrow H)$ at $\sqrt{s}=\unit[14]{TeV}$ ($\unit[8]{TeV}$). The combined signal strength $\mu$ for ATLAS \cite{ATLAS:2013mma} and CMS \cite{CMS:yva} is shown where we have assumed a SM-like BR. \label{fig:ggh}}
\end{figure}

The scalar $S$ can change the Higgs production rate in the gluon fusion channel due to $\lambda_{HS}$. We have calculated $\sigma(pp\rightarrow H)$ to the Next-to-leading (NLO) order with this additional scalar. We followed the calculation of Ref.~\cite{Bonciani:2007ex} and utilize the heavy scalar approximation. The MSTW2008 parametrization of parton density function (PDF) \cite{Martin:2009iq} implemented in LHAPDF \cite{Whalley:2005nh} has been used in our computation with the factorization scale $\mu_F$ and the renormalization scale $\mu_R$ set to be equal to $m_h$. We have utilized also the zero-width approximation for the Higgs boson to simplify the calculation and the resulting production cross section is shown in Fig.\ref{fig:ggh}. Since our model does not modify the branching ratio (BR) of the SM Higgs (the decay $H\rightarrow \gamma \gamma$ is modified with accidental symmetry, which we will discuss later on, but 
this loop induced decay is very small compared to the tree-level decays), we can compare the signal strength $\mu$ times $\sigma(pp\rightarrow H)_{\mathrm{SM}}$ measured by ATLAS \cite{ATLAS:2013mma} and CMS \cite{CMS:yva} to our model's prediction. The additional $S$ field decreases the Higgs gluon fusion production rate, with almost half the rate for large $\lambda_{HS}$ (small $\Lambda$) and small $m_S$. We obtain the suppression of $ggH$ production rate as opposed to the enhancement due to the negative sign of $\lambda_{HS}$.

\begin{figure}
\includegraphics[width=0.4\textwidth]{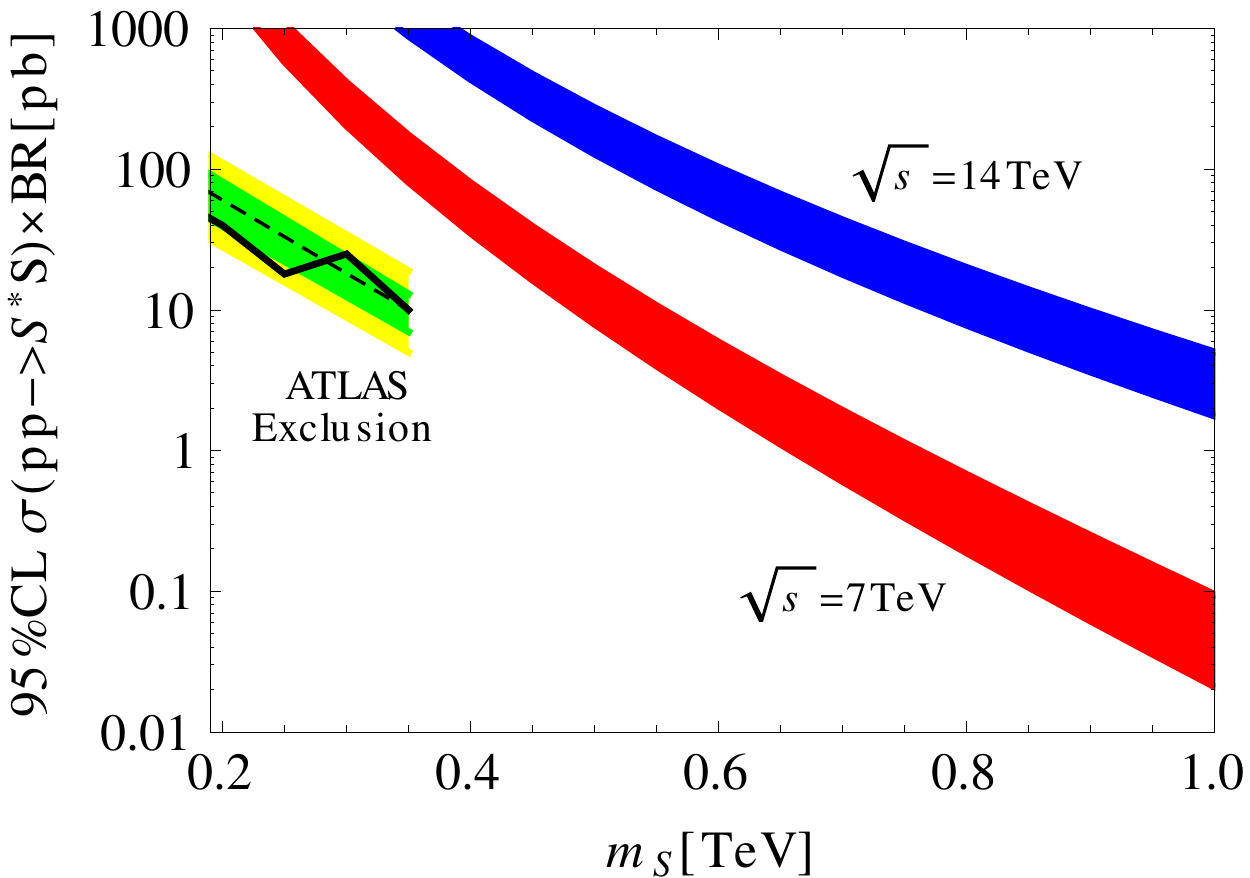}
\caption{The $S$ pair production cross section from gluon fusion channel is calculated for different value of $m_S$. The $95\%$ confidence level exclusion limit on $\sigma \times \mathrm{BR}$ for $\sqrt{s}=\unit[7]{TeV}$ by ATLAS is plotted. We assume $100\%$ BR of $\langle S^\dagger S \rangle$ into two jets. \label{fig:ggss}}
\end{figure}

The condensate $\langle S^\dagger S \rangle$ has to be heavier than the Higgs to trigger the EWSB, therefore it will decay to Higgs particles or two gluons. The scalar $S$ can be produced at the LHC, with the dominating production channel $gg\rightarrow S^*_i S_j$. The pair production of colored scalars with higher dimensional representation at LO in the gluon fusion channel has been calculated in Ref.~\cite{Martynov:2008wf,*Plehn:2008ae,*Bai:2010dj,*Idilbi:2010rs,*GoncalvesNetto:2012nt} and the result for our case is given in Fig.\ref{fig:ggss}. The resulting particles $S^*_i S_j$ will form two bound state pairs, with each pair decaying predominantly to $gg$ (2 jets) or to Higgs particles. Since the BR of $H\rightarrow b\bar{b}$ dominates, we would expect almost $70\%$ for $S^* S\rightarrow jjjj$ in the total cross section. The width of the band in Fig.\ref{fig:ggss} represents the factorization and renormalization scale dependence and the $\alpha_s$ uncertainty from RGE with extra $S$ contribution. 
In Fig.\ref{fig:ggss} we plot the ATLAS exclusion limit on pair production of new color scalar decaying to four jets \cite{ATLAS:2012ds}, where we have assumed $100\%$ BR to four jets. $m_S \lesssim \unit[350]{GeV}$ is excluded at $95\%$ confidence level and serves as our lower bound on $m_S$. Combining this result with the upper bound due to the triviality constraint above, the mass parameter of this model is very constrained, i.e.
\begin{align}
\unit[350]{GeV} \lesssim m_S \lesssim \unit[3]{TeV}.
\end{align}

\begin{figure}
\includegraphics[width=0.4\textwidth]{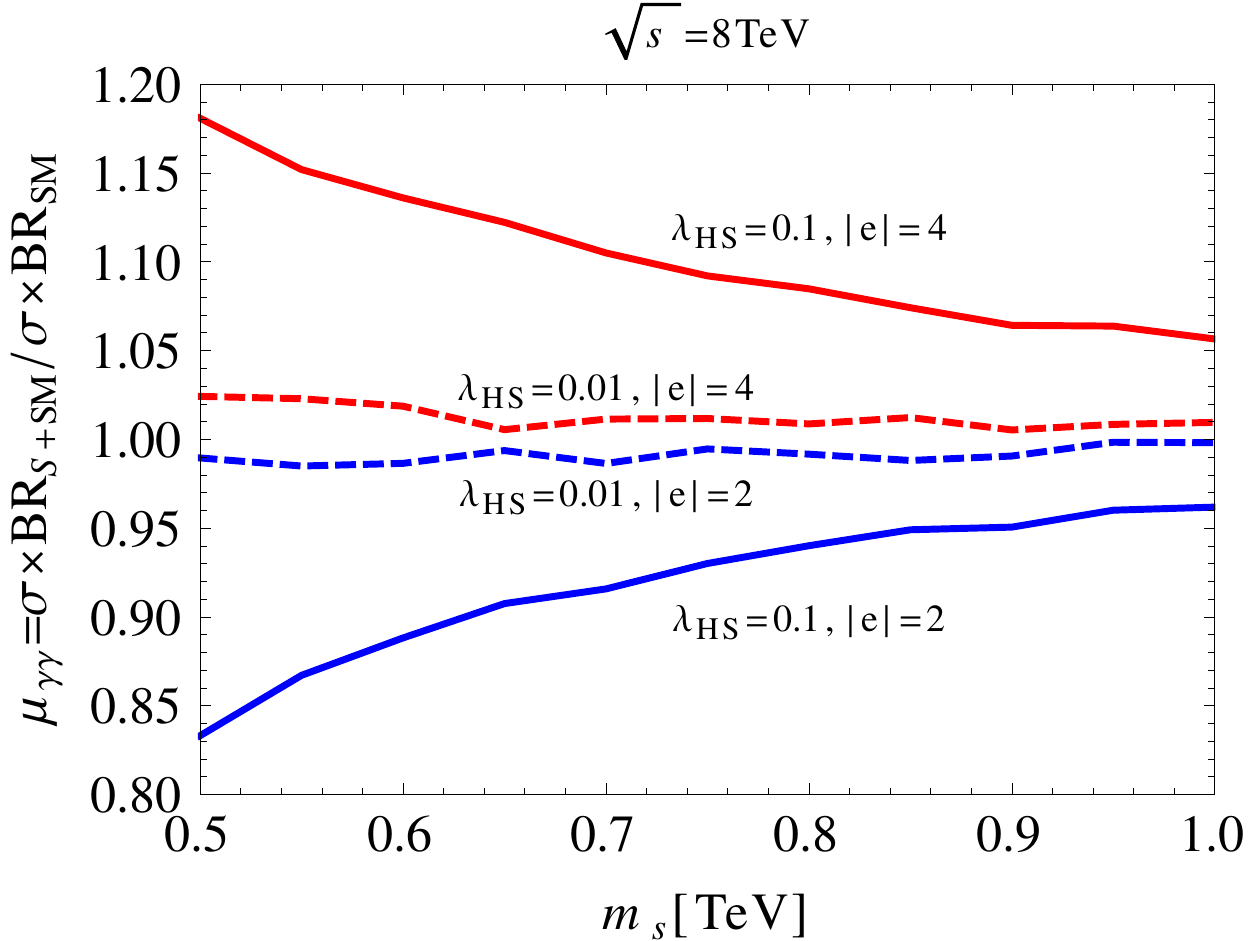}
\caption{The signal strength of $H\rightarrow \gamma \gamma$ branching ratio with the additional $S$ contribution relative to the SM prediction are plotted for different values of electric charge $e$ and $\lambda_{HS}$ of $S$. The large electric charge has to compensate the suppression of production cross section for $\mu_{\gamma \gamma}$ enhancement.  \label{fig:hgamma}}
\end{figure}

The $S$ field in Eq.~\eqref{eq:lagrangian} possesses an accidental $\mathrm{U}(1)$ symmetry due to the absence of the cubic term as we have imposed classical scale invariance in the Lagrangian. A priori this $\mathrm{U}(1)$ is another global symmetry, but if it is identified with the local $\mathrm{U}(1)_{Y}$ of the SM, we would obtain more interesting phenomenology. For instance the $H\rightarrow \gamma \gamma$ channel is enhanced by the additional $S$ running in the loop. Contrary to other scalar extension, enhancement of $H\rightarrow \gamma \gamma$ is obtained instead of suppression due to the minus sign of $\lambda_{HS}$ \cite{Carena:2012xa}. Strong enhancement of signal strength $\mu_{\gamma \gamma}$ for different values of $m_S$ can be obtained, with the result normalized to the SM prediction shown in Fig.\ref{fig:hgamma}. The signal strength $\mu_{\gamma\gamma}$ can be only enhanced by increasing the electric charge or $\lambda_{HS}$ of $S$ to compensate the suppression of production cross section. 
Compared to $\mu_{\gamma\gamma}\approx 1.65$ ($0.77$) reported by ATLAS \cite{ATLAS-CONF-2013-012} (CMS \cite{CMS:yva}) with the average $\mu_{\gamma\gamma}\approx 1.21$, our model would require large electric charge to explain the large $H\rightarrow \gamma \gamma$ anomaly. The large electric charge provides a possible alternative to study the $S$ particle via Drell-Yan production in linear collider.     

\section{Confinement of strongly coupled scalar field}
So far we have discussed the perturbative sector of the colored scalar $S$. We restricted the non-perturbative aspect of the model to the upscaling of the gap equation in Eq.~\eqref{eq:c2as}. Let us discuss a bit the physics in Eq.~\eqref{eq:c2as}. An analytical way to understand confinement in the quarks sector of QCD is to calculate the scaling of the gap equation from Dyson-Schwinger equation (DSE)
\begin{align}
\parbox{15mm}{\begin{fmffile}{dress}
\begin{fmfgraph}(40,25)
	        \fmfleft{g1}
		\fmfright{g2}
		\fmf{dashes}{g1,v1}
		\fmf{dashes}{g2,v1}
		\fmfdot{v1}
\end{fmfgraph}
\end{fmffile}}^{-1}&= \;
\parbox{15mm}{\begin{fmffile}{tree}
\begin{fmfgraph}(40,25)
	        \fmfleft{g1}
		\fmfright{g2}
		\fmf{dashes}{g1,g2}
\end{fmfgraph}
\end{fmffile}}^{-1}+\;
\parbox{15mm}{\begin{fmffile}{vertex}
\begin{fmfgraph}(40,25)
	        \fmfleft{v1}
		\fmfright{v2}
		\fmf{dashes}{v1,v3,v2}
		\fmf{curly,left,tension=0.2}{v1,v2}
		\fmfdot{v3}
		\fmfblob{4thick}{v2}
\end{fmfgraph}
\end{fmffile}}\;+...,
\end{align}
where we have utilized the rainbow-ladder approximation and only kept the leading order contribution to our analysis. The diagram above resembles the DSE for quark propagator, which can be solved within certain truncation scheme in order to obtain the critical value $X$ in 
\begin{align}
C_2(S)\alpha(\Lambda)\gtrsim X,
\label{eq:X}
\end{align}
for confinement to take place. However there are subtleties that one has to be careful when trying to extract the exact bound of $X$. First the value $X$ is gauge and truncation scheme dependent. Different values ranging from $0.6$ to $\pi/3$ have been obtained \cite{Miransky:1984ef,Alkofer:2000wg,Gribov:1999ui}. Lowering $X$ will allow us to consider lower representation of $S$ but in our analysis above we assume the conservative bound $X>0.8$. Second, the DSE for quark is linearizable with its linear form a Fredholm integral equation \cite{Miransky:1984ef,Cohen:1988sq,Leung:1989hw} as the wave function renormalization part and the self-energy part can be dealt separately in Landau gauge. However such privilege is not enjoyed by the scalar DSE as the integral equation
\begin{align}
F(p^2)=&p^2+\frac{3C_2\alpha_s}{4\pi}\left[ \int_0^{p^2}dq^2\frac{q^4}{p^2F(q^2)}+\int_{p^2}^\infty dq^2\frac{p^2}{F(q^2)}\right],
\end{align}
is not linearizable, where we have denoted the function $F(k^2)=Z(k^2)k^2+\Sigma^2(k^2)$. The main reason for such difficulty is due to the lack of confinement order parameter for scalar QCD. Comparing to fermionic QCD, the order parameter for confinement can be related to the degree of chiral symmetry breaking. From the perturbative calculation of the anomalous dimension of operator $\langle \bar{\psi} \psi\rangle$ and $\langle S^\dagger S \rangle$ in the same representation, it can be shown that 
\begin{align}
\gamma_{\langle \bar{\psi} \psi\rangle}=\gamma_{\langle S^\dagger S \rangle}+\mathcal{O}(\lambda_S).
\end{align}
Hence one can conjecture that the relevant order parameter $C_2\alpha_s$ at leading order for determining confinement should be the same for both fermionic QCD and scalar QCD, which we have assumed. In fact it has been argued that the scaling property for scalar and quark propagator in the infrared is identical \cite{Fister:2010yw}. This result can be verified in lattice QCD. 

Note that the QCD coupling becomes non-perturbative in the TeV regime even though the coupling is pretty small. This stems from the large value of $C_2$ which is responsible for condensation. As a consequence, the exact evolution of $\alpha_s$ cannot be precisely calculated in the TeV regime. However the coupling may become perturbative again at sufficiently small $\alpha_s$ or high energy. A similar conclusion was made in Ref.~\cite{Lust:1985aw}. Measuring $\alpha_s$ at high energy will provide an independent test for our model.

\section{Conclusion}
With no signature of any SM extension at the LHC and in other searches, the notion of naturalness deserves to be reexamined and other ideas of explaining the EW scale should be considered. We discussed in this letter a scenario where conformal symmetry plays an essential role and where the EW scale is a consequence of quantum effects. The idea of mass scale generation from a quantum effect, so called dimensional transmutation, is already implemented in the QCD sector of the SM. We have shown that it is possible to extend the success of QCD and to explain the existence of the EW scale by including a new scalar particle  
belonging to $\bm{15'}$ of $\SU3_c$. The extension is rather minimal and moreover unique if $X$ in Eq.~\eqref{eq:X} is greater than $0.8$.
The mass of this new colored boson is constrained such that it can be explored or ruled out by the LHC. The signature of this colored scalar field is comparatively clean. 
%Collider phenomenologies of diverse bound states which involve $S$, on the other hand, have yet to be studied. In principle it is also possible that bound states of $S$ with multiple quarks like $\langle qqqqS^* \rangle$ are expected to exist and could be produced at the LHC. Note, however, that $S$ must be pair produced if the accidental symmetry is not broken. We assume therefore that the production of these boundstates is a sub-leading effect. 
The accidental $\mathrm{U}(1)$ symmetry can also be probed in the $H\rightarrow \gamma \gamma$ signal strength if it is identified with the $\mathrm{U}(1)_Y$ of the SM. 
Furthermore, with a non-zero hypercharge the new colored boson
can be directly produced at linear colliders,
which will be our next target to investigate.
We leave the more detailed investigation of non-perturbative aspect of this model and the implication of EW phase transition to future work.

\vspace{0.5cm}
\noindent{\bf Acknowledgements:} 
We would like to thank Martin Holthausen, Julian Heeck and Tilman Plehn for useful discussions. K.~S.~L. acknowledges support by the International Max Planck Research School for Precision Tests of Fundamental Symmetries.

\bibliographystyle{apsrev4-1}
\bibliography{colorqcd}
\end{document}